# Consensus statement on the credibility assessment of ML predictors

Alessandra Aldieri[*], Dept. of Mechanical and Aerospace Engineering, Politecnico di Torino, Italy

Thiranja Prasad Babarenda Gamage, Auckland Bioengineering Institute, University of Auckland

Antonino Amedeo La Mattina, Medical Technology Lab, IRCCS Istituto Ortopedico Rizzoli, Italy

Yi Li National Clinical Research Center for Aging and Medicine, Huashan Hospital, Fudan University, Shanghai, China

Axel Loewe, Karlsruhe Institute of Technology (KIT), Karlsruhe, Germany

Francesco Pappalardo[**], Dept. of Drug and Health Sciences, University of Catania, Italy

Marco Viceconti, Dept. Industrial Engineering, Alma Mater Studiorum – University of Bologna, Italy.

[*]Authors are listed in alphabetical order; being this a consensus paper, all authors should be considered first authors with equal roles.

[**]Corresponding author. University of Catania, Department of Drug and Health Sciences, V.le A. Doria, 6, 95125 Catania (IT) – francesco.pappalardo@unict.it

*Abstract*

The rapid integration of machine learning (ML) predictors into in silico medicine has revolutionized the estimation of quantities of interest (QIs) that are otherwise challenging to measure directly. However, the credibility of these predictors is critical, especially when they inform high-stakes healthcare decisions. This position paper presents a consensus statement developed by experts within the In Silico World Community of Practice. We outline twelve key statements forming the theoretical foundation for evaluating the credibility of ML predictors, emphasizing the necessity of causal knowledge, rigorous error quantification, and robustness to biases. By comparing ML predictors with biophysical models, we highlight unique challenges associated with implicit causal knowledge and propose strategies to ensure reliability and applicability. Our recommendations aim to guide researchers, developers, and regulators in the rigorous assessment and deployment of ML predictors in clinical and biomedical contexts.

## Introduction

The advent of machine learning has ushered in a new era in biomedical engineering and in silico medicine, offering unprecedented capabilities in modelling complex biological systems and predicting clinically relevant outcomes. **Machine learning (ML) predictors** have emerged as powerful tools for estimating **quantities of interest (QIs)** that are difficult or impossible to measure directly, such as disease risk, treatment efficacy, or physiological parameters within the human body [1–7]

These advancements promise to transform healthcare by enabling personalised medicine, optimising therapeutic strategies, and improving patient outcomes. However, with great potential comes significant responsibility. The credibility of ML predictors is paramount, as inaccurate or unreliable predictions can lead to misdiagnosis, inappropriate treatments, and harm to patients [8–10].

The reliance on data-driven models introduces unique challenges. ML predictors often function as "black boxes," lacking transparency in how inputs are transformed into outputs. They depend heavily on the quality and representativeness of the training data and may inadvertently capture biases or spurious correlations. Furthermore, the absence of explicit causal relationships complicates the validation and regulatory approval processes.

Recognising the critical need for a standardised approach to the assessment of ML predictors' credibility, the **In Silico World Community of Practice** (https://insilicoworld.slack.com/, hosting 747 experts in In Silico Medicine; 35 joined the #credibility_machine_learning channel and participated in the consensus process) initiated a consensus-building process. This collaborative effort involved experts from various fields, including computational modelling, data science, clinical practice, and regulatory affairs. The goal was to establish a comprehensive framework that addresses theoretical underpinnings, methodological considerations, and practical steps necessary for rigorous credibility assessment. The manuscript is authored by the seven experts who materially contributed to its writing; however, all 35 experts supported the consensus statement reported here.

This position paper consolidates their insights into twelve detailed statements, forming a theoretical foundation and practical methodology for evaluating ML predictors in in silico medicine. Our focus centres on scenarios where the QI is a scalar quantitative value, although the principles are extensible to categorical or multivariate data. By contrasting ML predictors, which derive from implicit causal knowledge embedded in data, with biophysical predictors based on explicit scientific principles, we aim to illuminate the unique challenges and considerations associated with ML models.

Through an enhanced discussion of these statements, we delve into the complexities of credibility assessment, emphasising the importance of causal understanding, error decomposition, and strategies to ensure robustness and applicability. Ultimately, we aspire to provide a roadmap for researchers, developers, and regulators to navigate the intricate landscape of ML predictors, fostering their responsible and effective deployment in healthcare.

As common in consensus processes, the 12 statements reported below reflect (and to some extent incorporate) the current debate on the credibility and reliability of machine learning predictors in medicine. Thus, some of the concepts below resonate with ideas previously exposed in regulatory guidelines and reports; however, we propose that their organic structure, in a systematic assembly, constitutes an original contribution.

## Statements

**Statement #001**: There is a System of Interest (SI) whose internal state varies over time and space. $\Omega$ is the class of all quantities we can observe experimentally over SI; some of these quantities can be quantified "easily", whereas, for others, we can get a quantification only under particular conditions. One of these quantities is difficult to quantify, and thus, we would like to predict it by knowing the values of other quantities in $\Omega$ that are easier to quantify. We call this particular one Quantity of Interest (QI).

**Statement #002**: We subscribe to the *DIKW hierarchy* [11] as a representation of the process that links raw data to actionable knowledge. In this representation, we define *data* as the results of observing the SI. In particular, data can be *quantitative* or *categorical*. We generate categorical data through controlled experiments by observing the SI and assigning the QI to one of the pre-defined categories. We generate quantitative data through controlled experiments by measuring the QI of the SI under specific conditions. Raw data are of little use; they become *information* when annotated with all the additional data (metadata) necessary to define the context in which the data were observed. Typically, these are the answers to the questions "who", "what", "where", and "when". Information can be uplifted into tentative *knowledge*. For the purpose of this document, we define knowledge as a causation hypothesis that, if true, allows us to predict new data from observed ones. Knowledge becomes actionable (i.e., it can be used) when the causation hypothesis has resisted sufficient falsification attempts to be considered reliable; in this case, knowledge is uplifted into *wisdom*.

**Statement #003**: in the following, for clarity of exposition, we will focus on the special case where the QI is quantitative data and is a scalar. The generalisation of these reflections to the case where QI is categorical or multiscalar quantitative may pose operational but not conceptual challenges.

**Statement #004**: We must have some causal knowledge about the SI to predict the value the QI will assume in SI at a specific time and space. This causal knowledge can be *explicit* or *implicit*. *Explicit knowledge* is obtained through the scientific method: someone originally formulated that causal relationship that links the QI to other observable quantities as a hypothesis, and so far, multiple attempts failed to falsify it. Of course, the more numerous and extensive the falsification attempts, the more reliable this explicit knowledge is. Explicit knowledge is provided by the laws of physics and chemistry and the most investigated aspects of human physiology and pathology. Predictors built using explicit knowledge are called first-principle models, knowledge-driven models, mechanistic models, etc. Here, we will refer to them using the term *biophysical predictors*. *Implicit knowledge* is the causal knowledge that might be hidden inside a large data set (training set) observed experimentally over the SI. From these data, we can observe a correlation between specific observable quantities and the QI. However, we can develop confidence that it is indeed causation only by extensively attempting falsification by induction with additional test sets. Predictors built using implicit knowledge are called data-driven models, phenomenological models, etc. Here, we will refer to them using the term *machine learning (ML) predictors*.

**Statement #006**: We define the *credibility* of a predictor as the lowest accuracy with which the predictor provides an estimate of the QI over the entire information space that represents all possible states of the SI. This definition is operationally impossible to achieve, as we would need a true value of the QI for each possible internal state of the SI to compute it, which would make the predictor useless. All credibility frameworks (metrological credibility of measurements, statistical credibility of inference, Verification, Validation and Uncertainty Quantification of knowledge-based predictors) use a similar approach to estimate credibility. They decompose the prediction error calculated over a relatively small number of true values of QI into the possible sources of error (which depend on the nature of the predictor) and control that their distribution is as expected for that type of error. For example, the aleatoric component of the prediction error should be distributed normally. Suppose the estimate of the predictor error is acceptable and its components are distributed as expected. In that case, we can assume the predictor is well-behaved and accepts the induction risk of estimating the credibility with a relatively small number of true values.

**Statement #007**: Generally, the quantities we can observe for the system of interest are not mutually independent. Thus, the particular values we can observe for the QI correlate to the values assumed by other observable quantities $I \in \mathcal{I}$. Ideally, we would like to solve the *causation problem* and determine which observable quantities are necessary and sufficient to define the value of the QI. But hereinafter, we will only assume we know $C \in \mathcal{I}$. These observable quantities are sufficient but not necessarily necessary to define the value of the QI. On the contrary, we will not discuss here the case where some of the necessary quantities are not part of the observable set $\mathcal{I}$. In other words, we will assume that every quantity necessary to cause the QI is observable, although not necessarily known.

**Statement #008**: The QI usually correlates with each observable quantity only for a finite range of possible values of such quantity; we call it *limits of validity*. Hereinafter, we assume that we know the limits of validity for the QI and for each observable quantity that correlates with the QI. If $C$ includes only two quantities, we could, in principle, plot all true values $O$ that the QI can assume as a 3D surface (2-manifold), but because all quantities in $C$ are limited, that would only be a *patch*, a portion of that surface. If $C$ includes $n$ quantities, that entire *information space* can be represented as a bounded n-manifold.

**Statement #010**: The *credibility* of a predictor is the knowledge of the error affecting the estimation of the outputs for any possible value of the inputs. True credibility can never be achieved since this would require infinite validation experiments. However, credibility itself can be estimated using the same general process for both biophysical and ML predictors:

- *S1. Context of use and error threshold*
- The first step is to define the context of use for that predictor. In particular, it is necessary to define the maximum error $\underline{\varepsilon}$ affecting the information that still makes it useful for that use. The first requirement that any predictor of the QI must satisfy to be considered credible is that in all points of the solution space where the predictor accuracy is tested, the error is $\varepsilon_e < \underline{\varepsilon}$.
- *S2. Source of true values*

The second step is to define the source of the true values. We postulate that true values can be obtained only through measurement, using a measurement chain that ensures both for the QI and the correlated quantities **C** a class of accuracy that is at least one order of magnitude smaller than the maximum error defined for a context of use of the predictor.

- *S3. Quantification of prediction error*

The predictor's error can be quantified by sampling the solution space through controlled experiments in which the correlated quantities ***C*** are imposed or measured, and the true values for the QoI and the corresponding correlated quantities are quantified.

- *S4. Identification of the sources of error*

Depending on the nature of the predictor, we now identify the possible sources of such prediction error. This is the most delicate step and requires a profound understanding of how each particular class of predictors operates.

- *S5. Decomposition of the prediction error*

This step is even more challenging: once the various sources of predictor error are identified, we must find ways to estimate how the overall prediction error can be divided among these sources. In some cases, this might require generating true values under special conditions where all but one source of error is excluded.

- *S6. Critical review of error distributions*

As we identify the various potential sources of error affecting the predictor, we also can define some expectations of how that component of the prediction error should be distributed over repeated predictions across the solution space. Once the prediction error is decomposed over its sources, we check if these expectations are confirmed.

- *S7. Robustness to biases and applicability*

Once the predictor is considered credible for a context of use, it will be used routinely; the last step in the credibility assessment is to infer if there might be in this routine use additional biases, which may alter the accuracy of the predictor, that do not appear in the controlled experiments used to quantify the prediction error. A special case is that of applicability: we need to ensure that when the predictor is used routinely, the QoI and its correlates never exceed the limits of validity we defined during the credibility assessment, and if they do, how the credibility of the predictor may be compromised.

**Statement #011**: we can now consider if there is any difference between biophysical and ML predictors when these seven steps to assess credibility are applied. Most of the steps are identical for the two types of predictors, except these:

- *S1. Identification of the sources of error*

For biophysical models, there is a consensus that the three sources of error are *numerical, aleatoric*, and *epistemic* uncertainty. Numerical uncertainty is due to the computational methods used to solve the mathematical forms used to express the explicit knowledge on which the model was built. Aleatoric uncertainty is caused by the measurement uncertainty affecting the input values. Epistemic uncertainty is caused by the errors associated with applying that explicit knowledge in the prediction of the QI. ML predictors are affected by aleatoric uncertainty and may be affected by some forms of numerical uncertainty. While there is no explicit knowledge of an ML predictor that may cause epistemic uncertainty, the choice of the model form used to train the predictor is in itself a form of epistemic uncertainty. Thus, we can conclude that both biophysical and ML predictors are affected by similar sources of error.

- *S2. Decomposition of the prediction error*

For biophysical models, such decomposition is achieved using the verification, validation, and uncertainty quantification process. Here, we postulate that similar processes can also be devised for ML predictors.

- *S3. Robustness to biases and applicability*

This is where biophysical predictors diverge from ML predictors. The explicit knowledge used to build a biophysical predictor should ensure that the input set is necessary and sufficient to predict the QI; in other words, no other quantity observable in the system of interest and not included in the input set can significantly affect the QI. In this case, the problem of robustness to biases is reduced to that of applicability, which in turn merely requires that the prediction error varies smoothly over the input space. This is a reasonable assumption for biophysical models, as far as the inputs are not too close to the limits of validity of the explicit knowledge used to build them. However, the input set is *sufficient* for ML predictors but not necessarily necessary to predict the QI; on the one hand, the ML predictor may be overfitted, and the input sent includes more inputs than those required, which may cause a loss of robustness in the predictor; on the other hand, for particular reasons, there might be an observable quantity that generally affects QI. One example is when, in the particular training set used to develop the ML predictor, an observable quantity was "silent", for example, because it did not vary too much in that particular set of experimental observations. As a result, that quantity is not included in the input set, even if it should have been.

This last point is the most critical. Because we are not sure if the input set of the predictor is necessary, we cannot reduce the issue of robustness of biases to the mere applicability assessment.

**Statement #012**: to ensure robustness to biases, developers of ML predictors have two strategies. The first, already mentioned by the FDA, is the Total Product Life Cycle (TPLC) approach. The ML predictor is certified for use in cases with characteristics similar to the test sets used for the credibility assessment. As new, more diverse test sets are added, the context of use can be broadened. The second is to add to the ML predictor a safety layer, which collects as many observable values as possible for the subject target of the ML prediction and compares them to the distribution of observable values in the training and test set cohorts. Suppose the subject's data suggest he/she is not a member of the training/test population. In that case, the ML system refuses to make the prediction or warns the clinical user that the prediction might be inaccurate. The main issue with this approach is that even if the correlation analysis suggests the ML predictor can be built, say with only five out of 20 observable quantities, all twenty values should be collected and stored for every subject in the training and test sets to make this safety layer possible.

# Discussion

*Understanding the System of Interest and Quantity of Interest*

In the realm of in silico medicine, we grapple with complex and dynamic systems—human physiology being a quintessential example—where countless variables interact intricately over time and space. Within such a **System of Interest (SI)**, identifying and predicting specific **Quantities of Interest (QIs)** is crucial for advancing medical understanding, diagnostics, and treatments. However, direct measurement of these QIs is often impractical due to technical limitations, invasiveness, or ethical considerations [10,12,13].

For instance, predicting cardiac electrical activity across heart tissue is vital for diagnosing arrhythmias but requires invasive procedures. Leveraging ML predictors to estimate this QI from non-invasive measurements like electrocardiograms exemplifies the broader challenge: accurately predicting difficult-to-measure QIs using accessible data [14].

ML predictors offer a compelling solution by harnessing patterns within large datasets to infer the QI based on other observable quantities in $\Omega$. However, the credibility of these predictions hinges on the robustness of the model and the quality of the underlying data.

*The DIKW Hierarchy in Credibility Assessment*

The transformation from raw data to actionable wisdom is foundational in developing credible predictors. The **Data-Information-Knowledge-Wisdom (DIKW) hierarchy** provides a structured pathway:

- **Data**: Raw observations lacking context or meaning.
- **Information**: Data enriched with metadata, providing context and facilitating interpretation.
- **Knowledge**: Formulation of causation hypotheses that explain relationships within the information.
- **Wisdom**: Knowledge validated through rigorous testing, deemed reliable for decision-making.

Moving up the DIKW hierarchy in ML predictors requires meticulous data curation, ensuring that the information accurately reflects the SI's complexity. Establishing knowledge involves identifying genuine causal relationships rather than mere correlations, a significant challenge when dealing with implicit knowledge.

For example, in developing an ML predictor for tumour growth, raw imaging data (data) are annotated with patient information and imaging parameters (information). Hypotheses about how specific features influence growth rates are formulated (knowledge) and, after extensive validation, inform treatment decisions (wisdom).

*Causal Knowledge: Explicit vs. Implicit*

The distinction between **explicit** and **implicit** causal knowledge is pivotal in understanding the credibility challenges of ML predictors.

**Biophysical Predictors** rely on explicit knowledge derived from well-established scientific laws. They model systems using equations and principles that describe known causal mechanisms. For instance, modelling the biomechanics of bone fracture healing using finite element analysis depends on explicit laws of mechanics and biology.

This explicitness affords transparency and interpretability. Inputs, outputs, and their relationships are clearly defined, facilitating validation and error analysis. However, biophysical models may struggle with complexity and computational demands when scaling to highly intricate systems.

**ML Predictors**, conversely, operate on implicit knowledge extracted from data. They identify patterns and associations without explicit causal models. While this allows them to handle complex, high-dimensional data, it also means that the causal pathways are not inherently understood.

For instance, an ML model predicting patient outcomes based on electronic health records may uncover associations between certain lab values and mortality risk. Without explicit causal understanding, it's challenging to discern whether these associations are genuine or artifacts of the data.

The reliance on implicit knowledge raises concerns about the model's ability to generalize beyond the training data, especially when encountering new or rare scenarios not represented in the original dataset.

Given the respective strengths and limitations of these approaches, the integration of explicit and implicit knowledge through hybrid modeling represents a promising avenue. Hybrid methods aim to combine the interpretability and theoretical grounding of biophysical models with the data-driven flexibility of ML.

Such integration can take several forms:

- Physics-Informed Machine Learning (PIML): In this approach, physical laws and constraints derived from biophysical models are incorporated into the ML training process. This integration can improve the generalizability and plausibility of ML predictors while reducing the risk of spurious associations.
- Model-Agnostic Hybrid Frameworks: These frameworks allow biophysical models and ML predictors to work in tandem. For instance, ML models can augment biophysical simulations by predicting parameters or initial conditions, while biophysical models provide the mechanistic foundation to guide ML predictions in areas with limited data.
- Sequential or Parallel Hybrid Systems: Biophysical and ML models can operate sequentially or in parallel, with outputs from one feeding into the other. For example, an ML model may predict outcomes that are then validated or adjusted using biophysical simulations to ensure consistency with known causal mechanisms.

The integration of these approaches is particularly relevant in biomedical contexts where explicit knowledge is available only for portions of the problem. For example, Karniadakis et al. [15] highlight the potential of hybrid methods to address challenges in modelling complex systems, such as blood flow or tumour growth, where certain physical laws are well understood, but high-dimensional and stochastic data play a significant role. By leveraging the complementary strengths of biophysical and ML models, hybrid approaches have the potential to enhance predictive performance, improve interpretability, and increase the credibility of ML-based solutions in biomedical applications.

*Defining and Estimating Credibility*

Assessing the credibility of a predictor involves estimating its reliability across all possible states of the SI—a daunting task given the system's complexity. To make this feasible, we employ statistical sampling and error analysis.

**Quantifying Prediction Error**: By comparing the predictor's outputs to true QI values obtained from controlled experiments or high-precision measurements, we quantify the prediction error.

**Identifying Sources of Error**: Understanding where errors originate is crucial. Errors may stem from:

- **Numerical Uncertainty**: Approximation errors in computational methods.
- **Aleatoric Uncertainty**: Inherent variability or randomness in the data.
- **Epistemic Uncertainty**: Lack of knowledge or model limitations.

**Decomposing Errors**: Breaking down the total error into these components allows targeted improvements. For example, if numerical uncertainty is significant, refining computational algorithms may help. If epistemic uncertainty dominates, expanding the model's knowledge base or incorporating additional variables may be necessary.

**Analysing Error Distributions**: Errors should conform to expected statistical distributions. For instance, aleatoric errors are often assumed to be normally distributed. Deviations from expected patterns may indicate model biases or unaccounted factors.

By systematically assessing errors, we can infer the predictor's credibility and identify areas for enhancement.

*Challenges Unique to ML Predictors*

ML predictors present specific challenges that can impact credibility:

- **Overfitting**: ML models may learn noise or irrelevant patterns in the training data, leading to poor generalisation. Overfitting can be mitigated through techniques like cross-validation, regularisation, and pruning, but it remains a critical concern.
- **Biases in Training Data**: The model inherits biases present in the training data. If certain populations or conditions are underrepresented, the predictor may perform inadequately for those groups. Addressing this requires careful data collection and augmentation strategies.
- **Missing Necessary Inputs**: ML models may omit variables that are causally relevant to the QI but were not included in the training data. This can result in models that perform well on training data but fail when these variables vary in new data.
- **Interpretability and Transparency**: Many ML models, especially deep learning architectures, lack interpretability. This "black box" nature hinders understanding of how predictions are made, complicating error analysis and validation.
- **Data Quality and Consistency**: ML models are sensitive to data quality. Inconsistent data collection methods, measurement errors, and missing values can adversely affect model performance.
- **Dynamic Systems and Temporal Changes**: Biological systems are dynamic, and relationships between variables may change over time. ML models trained on historical data may not account for temporal shifts, reducing credibility in current predictions.

**Strategies for Enhancing ML Predictor Credibility**

To address these challenges, we propose the following strategies:

*Total Product Life Cycle (TPLC) Approach*

The TPLC approach emphasises continuous evaluation and improvement:

- **Lifecycle Monitoring**: Monitor the predictor's performance throughout its deployment, not just during initial validation.
- **Data Integration**: Incorporate new data as it becomes available, retraining the model to capture emerging patterns and address changes in the SI.
- **Feedback Mechanisms**: Implement systems for users to report discrepancies or errors, facilitating iterative refinement.
- **Regulatory Alignment**: Ensure that updates and changes comply with regulatory requirements, maintaining transparency and documentation.

This approach acknowledges that ML predictors operate in evolving environments and must adapt to maintain credibility.

*Implementation of a Safety Layer*

A safety layer enhances robustness by:

- **Assessing Input Validity**: Before making predictions, the model evaluates whether input data falls within the scope of its training data.
- **Outlier Detection**: Identifying and handling inputs that are significantly different from the training data helps prevent unreliable predictions.
- **User Alerts and Warnings**: Providing confidence scores or warnings when predictions may be less reliable informs users of potential risks.
- **Fallback Mechanisms**: In cases of high uncertainty, the model may defer to human judgment or alternative methods.

Implementing a safety layer requires comprehensive data collection, including variables not directly used in predictions but essential for assessing input validity.

**Comprehensive Data Collection**: Collecting and storing all relevant observable quantities for each subject in the training and test sets enables effective applicability analysis. Even if the correlation analysis suggests that only a subset of variables is needed for prediction, retaining additional data supports robustness checks.

**Position the consensus statement within the current debate**

Most related literature on machine learning in medicine focuses on three concepts: interpretability [16], explainability [17], and reliability [12]. *Interpretability* refers to the degree to which a human can understand the cause-and-effect relationship between the inputs to a machine learning model and its outputs; *Explainability* refers to the ability to provide a clear and meaningful description of how a machine learning model arrives at its predictions, typically in terms understandable to humans; *Reliability* refers to the consistency and dependability of a machine learning model's predictions when exposed to varying input conditions, including unseen data or scenarios.

Here we focus on extending the concept of *credibility* used in regulatory science for first-principle (biophysical) predictive models (see, for example (see, for example, [18][19]) to machine learning predictors. Credibility assessment has a more stringent objective than the reliability assessment: while reliability aims to estimate the probability that in one case, the predictor may be unacceptably wrong, credibility in contrast uses verification validation, uncertainty quantification and applicability analysis to ensure that in no case (within the limits of validity of the predictor) the predictive error will exceed the acceptability threshold defined in the relation to the specific clinical context of use of the predictor.

Interpretability and explainability are surely desirable in an ML predictor, but they will not tell us if and when the predictions are unacceptably wrong. Reliability is well suited for industrial applications of ML, but in the medical regulatory science, credibility assessment is required.

In light of the recent FDA draft guidance on the use of artificial intelligence to support regulatory decision-making for drug and biological products[1], we reviewed its recommendations in the context of our proposed framework. The draft guidance emphasizes a risk-based credibility assessment tailored to the context of use, aligning well with the principles outlined in our manuscript. Notably, both the FDA guidance and our framework recognize the role of the context of use in defining acceptability thresholds and validation plans. While the FDA's focus on risk assessment as a foundation for credibility contrasts with our emphasis on prediction error thresholds, this difference appears complementary. Additionally, the guidance reiterates that acceptability thresholds should emerge from the context of use, a principle we explicitly discuss. This comparison highlights areas of convergence and potential complementarity between the two approaches

## Conclusions and Recommendations

The credibility of ML predictors is of paramount importance in *in silico* medicine, given their increasing role in informing clinical decisions. While ML models offer significant advantages in handling complex data and uncovering patterns, they introduce unique challenges that must be meticulously addressed to ensure reliability and trustworthiness.

**Key Conclusions**

- **Implicit Knowledge Limitations**: ML predictors' reliance on implicit knowledge makes them susceptible to biases, overfitting, and issues arising from missing necessary inputs.
- **Error Decomposition is Crucial**: Understanding and quantifying different sources of error are essential for improving model performance and credibility.
- **Robustness Strategies are Necessary**: Implementing strategies like the TPLC approach and safety layers enhances the predictor's robustness and applicability across diverse scenarios.

**Recommendations**

1. **Develop Standardized Assessment Protocols**: Establish guidelines tailored specifically for ML predictors, incorporating the seven-step credibility assessment process outlined in this consensus.
2. **Comprehensive Data Collection**: Encourage the collection of extensive, high-quality datasets, including all potentially relevant variables, to support robust model training, validation, and applicability assessments.
3. **Advance Validation Methods**: Invest in research to develop validation, verification, and uncertainty quantification methods adapted to the unique aspects of ML predictors.
4. **Enhance Model Transparency**: Promote the use of interpretable ML models or techniques that provide insights into model decisions, facilitating error analysis and building user trust.
5. **Regulatory Engagement**: Collaborate with regulatory agencies to align credibility assessment methods with compliance requirements, ensuring that ML predictors meet the necessary standards for clinical use.

---

[1] https://www.fda.gov/regulatory-information/search-fda-guidance-documents/considerations-use-artificial-intelligence-support-regulatory-decision-making-drug-and-biological

6. **Education and Training**: Implement educational initiatives to enhance understanding of ML predictor development, validation, and credibility assessment among researchers, clinicians, and stakeholders.

7. **Interdisciplinary Collaboration**: Foster collaboration among data scientists, domain experts, clinicians, and regulators to address the multifaceted challenges of ML predictor development and assessment.

8. **Continuous Monitoring and Updating**: Establish mechanisms for ongoing performance monitoring in real-world settings and update models as new data and insights emerge.

9. **Transparency in Model Development**: Maintain clear documentation of model architectures, training procedures, data sources, and limitations to facilitate critical evaluation and replication efforts.

10. **Ethical Considerations**: Incorporate ethical guidelines to address potential biases, fairness, and equity in ML predictor deployment, ensuring that models serve all populations effectively.

By adhering to these recommendations, the community can enhance the credibility of ML predictors, ensuring they provide accurate, reliable, and clinically valuable estimations of QIs. This will ultimately contribute to improved patient outcomes, foster trust in in silico methods, and advance the field of computational medicine.

## Acknowledgments

We acknowledge the role of Prof Marco Viceconti as chair of the consensus panel that produced this position paper. The In Silico World Community of Practice was supported by the European Commission through the H2020 project "In Silico World: Lowering barriers to ubiquitous adoption of In Silico Trials" (topic SC1-DTH-06-2020, grant ID 101016503).